\documentstyle[aps,multicol,epsf]{revtex}
\begin{document}
\draft
\title{Critical Percolation and Transport in Nearly One Dimension}
\author{A.~ N.~ Samukhin$^{1,2}$, V.~ N.~ Prigodin$^{1,3}$, and L.~Jastrab\'\i k$^2$}
\address{$^1$A. F. Ioffe Physical \& Technical Institute, 194021
St.Petersburg, Russia\\
$^2$Institute of Physics AS CR, Na Slovance 2, 180
40 Prague 8, Czech Republic\\
$^3$Max-Plank-Institute f\"ur Physik komplexer
Systeme, Au\ss enstelle Stuttgart, Heisenbergstr. 1, D-70569 Stuttgart,
Germany}
\maketitle

\begin{abstract}
  A random hopping on a fractal network with dimension slightly above one, $d
  = 1 + \epsilon$, is considered as a model of transport for conducting
  polymers with nonmetallic conductivity. Within the real space
  renormalization group method of Migdal and Kadanoff, the critical behavior
  near the percolation threshold is studied.  In contrast to a conventional
  regular expansion in $\epsilon$, the critical indices of correlation length,
  $\nu =\epsilon ^{-1}+O(e^{-1/\epsilon })$, and of conductivity, $t\simeq
  \epsilon^{-2}\exp (-1-1/\epsilon )$, are found to be nonanalytic functions
  of $\epsilon $ as $\epsilon \rightarrow 0$.  Distribution for conductivity
  of the critical cluster is obtained to be gaussian with the relative width
  $\sim \exp (-1/\epsilon )$. In case of variable range hopping an ``1-d
  Mott's law'' $\exp \left[ -\left( T_t/T\right) ^{1/2}\right] $ dependence
  was found for the DC conductivity.  It is shown, that the same type of
  strong temperature dependence is valid for the dielectric constant and the
  frequency-dependent conductivity, in agreement with experimental data for
  poorly conducting polymers.
\end{abstract}

\pacs{PACS numbers: 73.61.Ph, 71.30.+h, 72.90.+y, 05.60.+w}

\begin{multicols}{2}


Conducting polymers represent a large class of new materials with a great
variety of transport properties \cite{ts92}. The room-temperature conductivity
($\sigma _{RT}$) of some of them attains the metallic value and the
temperature and frequency dependencies of their conductivity are closely to be
metallic. The nature of the metallic phase is presently a subject of intensive
study \cite{icsm'94}. One point is that the metallic state is provided by
strong interchain coupling in these materials~\cite{nps89}. In polymers with
moderate $\sigma _{RT}$ (of the order of several hundreds $S/cm$) the
conductivity, as a rule, decreases with decreasing temperature. Because this
decay follows a power law in a large temperature interval, presumably these
materials are near the metal-insulator transition which happens at the
critical interchain coupling. Poorly conducting samples with $ \sigma _{RT}$
of the order or less 1 $S/cm$ demonstrate a behavior that can be classified as
a dielectric one \cite{wjrme90}: it is similar to that observed in amorphous
semiconductors. Namely, DC conductivity is strongly dependent on temperature
and its best fit is given by an ``1-d Mott law'': $\Sigma _{DC}\propto \exp
-(T_0/T)^{1/2}$.  For a variable range hopping mechanism of transport the
temperature dependence of conductivity is \cite{md79,shef84}: $\Sigma
_{DC}\propto \exp -(T_0/T)^{1/(d+1)}$, where $d>1$ is the system's dimension.
This is not true, however, in a 1d case \cite{kj73}, for which it should be the
Arhenius dependence, $\Sigma _{DC}\propto \exp -T_0/T$.

Experimental measurements of the microwave conductivity and the dielectric
constant of these poorly conducting samples \cite{wjrme90,joea94} revealed
that both are strongly dependent upon temperature too, most probably according
to the same 1-d Mott's law. The theory of hopping transport predicts, however,
only a very weak power temperature dependence for the frequency-dependent
conductivity and the dielectric constant in two-- and three--dimensional
systems~\cite{bb85}.

In the present Letter we suggest an explanation for these peculiar features of
conducting polymers that is based on the specific structure of the polymer's
network. In the stretched polyacetylene this network is formed by coupled
polymer chains oriented along some direction. Electronic micrographs shows
that in these substances polymeric chains are organized into {\em fibrils
  }\cite{ts92}, which may be distinctly seen to be subdivided into smaller
ones \cite{ar87}. In a non-fibrillar form of conducting polymers, like
polyaniline, X--rays data reveal the existence of highly ordered
``crystallinity regions'' with metallic properties \cite{ts92}.  Therefore the
whole network in the stretched polyaniline may be thought of as constructed
from long one--dimensional polymer chains randomly coupled by metallic
crystallinity islands of various sizes.  The volume fraction of latter ones
can be small.

We assume here that a polymer structure represents {\em nearly one-dimensional
  fractal}. That means a specific kind of polymer's chains organization,
defined in the following way: Choose a three-dimensional cube with the edge
$L$. Chains, which are coupled within this cube, form a set of bundles
disconnected from each other. If for large enough $L$ the cross--section of
the maximum bundle is proportional to $L^\epsilon $, where $0\leq \epsilon
\leq 2 $, then we shall call the system $d^{*}=1+\epsilon $ --dimensional.
Obviously $\epsilon =0$ for purely one--dimensional systems (sets of uncoupled
chains). Note, that if one assume chains to be connected either with low
concentration of uncorrelated interchain links, or with weak links (their
resistivities being high compared to intrachain ones in our example), then we
are dealing with a {\em \ quasi--one--dimensional} system~\cite{nps89}, which
is three--dimensional according to our definition.

The transport mechanism in conducting polymers in the localized phase is
assumed to be the variable range hopping type (VRH) \cite{jpmme94}. The
regular method to treat VRH models is the effective medium
approximation~\cite{bb85}, which gives wrong results in the nearly--1d case.
For example, for the percolation model it gives a threshold concentration of
broken bonds $c_t\approx \epsilon $, while $ c_t\approx \exp \left(
-1/\epsilon \right) $, as we shall see later. The results for the critical
exponents are also wrong in this case. We choose the following approach. We
will first study the nearly--1d percolation system exactly. The VRH model is
reduced to the percolation problem by constructing the effective percolation
lattice \cite{shef84,bb85}.

Our first aim is the study of the critical behavior of the conductivity near
the percolation threshold in a $d$--dimensional lattice, where $d$ is close to
the lower critical dimensionality, i.e. $d=1+\epsilon $, $\epsilon \ll 1$. The
real space renormalization group of Migdal and Kadanoff (RGMK) \cite
{mg76,kd76,tm96}, that is exact at $d=1$, expected to be the appropriate tool
if $d$ is close to one. This method was applied to the percolation
conductivity problem by Kirkpatrick \cite{k77}, but the case of dimensionality
close to $1$ did not get any special attention.

\begin{figure}[tbp]
\epsfxsize=8cm
\epsffile{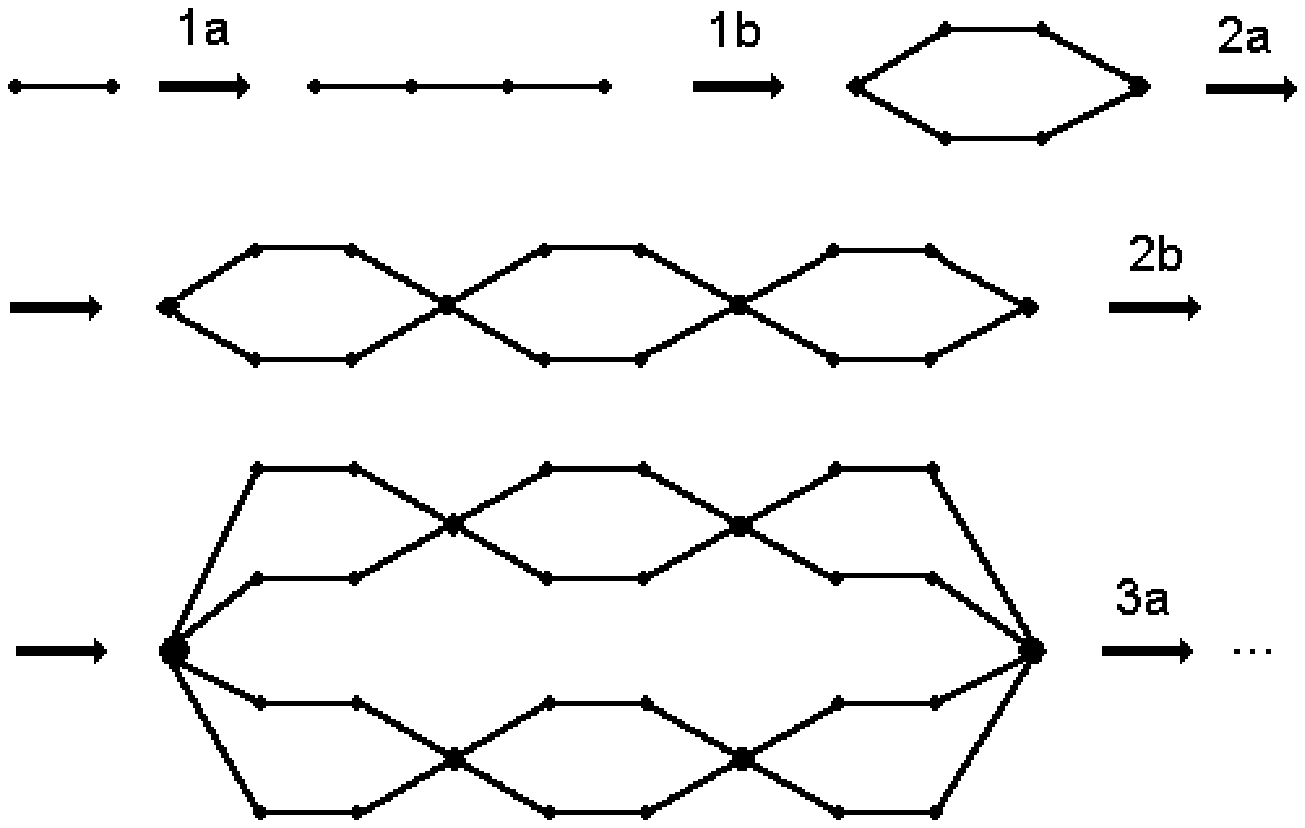}
\vspace{1cm}
\narrowtext
\caption{The two initial stages of $(2,3)$ hierarchical structure formation.}
\label{hierstr}
\end{figure}

The RGMK method may be formulated as follows: the $d$--dimensional hypercubic
lattice is replaced by a $(m,n)$ hierarchical structure (see below) with
$m=n^{d-1}$, and $ n\rightarrow 1$ afterwards. The $(m,n)$ hierarchical
pseudolattice may be constructed by the infinite repetition of two subsequent
steps, as illustrated in Fig.\ref{hierstr}:\\a) Formation of a $n$-length
chain from $n $ bonds;\\b) Formation of a $m$-bundle with $m$ chains in a
cross-section.\\ Bundles obtained by this way are used as elementary bonds at
the next stage, etc. To have a continuous RG transformation instead of discrete
one, one should proceed to infinitesimal transformation, setting $n=1+\nu
\,,m=1+\epsilon \nu ;\,\nu \rightarrow 0$, where $\,\epsilon =d-1$. At this
step we introduce the continuous scale variable $\lambda =\exp \left( \nu
l\right) $, where $l$ is the order of bonds in the hierarchical structure.

\vskip 0.5cm

It appears to be convenient to work with conductivity  $\Sigma $ and
resistivity $R=1/\Sigma $ distribution functions in the Laplace
representation:
\begin{equation}
P(\sigma )\equiv \langle \exp (-\sigma \Sigma )\rangle ;\quad Q(s)\equiv
\langle \exp (-sR)\rangle .  \label{distrdef}
\end{equation}
This is justified by the fact that for the chains or the bundles formation we
have simply sums of independent random resistivities or conductivities,
respectively.  These functions are related by the integral transformation:
\begin{equation}
P(\sigma )=1-\sqrt{\sigma }\int\limits_0^\infty \frac{ds}{\sqrt{s}}J_1(2%
\sqrt{s\sigma })Q(s),
\label{hankel}
\end{equation}
where $J_1(s)$ is the Bessel function.  For the hierarchical structures, we
arrive after the transition to a continuous transformation, at the following
equation:
\begin{eqnarray}
&&\lambda \frac{\partial Q(s,\lambda )}{\partial \lambda }+(1-\epsilon )%
\frac{\partial Q(s,\lambda )}{\partial s}-Q(s,\lambda )\ln Q(s,\lambda )
\nonumber \\
&+&\epsilon \sqrt{s}\int\limits_0^\infty \frac{d\sigma }{\sqrt{\sigma }}J_1(2%
\sqrt{s\sigma })P(\sigma ,\lambda )\ln P(\sigma ,\lambda )=0.  \label{main}
\end{eqnarray}

Finite concentration of broken bonds $c_b$ may be considered by introducing
the boundary conditions:
\begin{equation}
Q(s,\lambda )\mid _{s=0}=1-c_b\quad \Longleftrightarrow \quad P(\sigma
,\lambda )\mid _{\sigma =+\infty }=c_b.  \label{bb}
\end{equation}
Setting in Eq. (\ref{main}) $s\rightarrow 0$, we obtain the RGMK equations
for broken bonds concentration,
\begin{equation}
\lambda \frac{dc_b}{d\lambda }=\epsilon c_b\ln c_b-(1-c_b)\ln (1-c_b).
\label{rb}
\end{equation}
There are three fixed points, two stable ones, $c_b=0$ (regular lattice) and
$c_b=1$ (completely broken lattice), and one unstable fixed point $c_b=c_t$,
where the threshold concentration, $c_t$ ($0<c_t<1$) is given by
\begin{equation}
\epsilon c_t\ln c_t=(1-c_t)\ln (1-c_t).  \label{thres}
\end{equation}

\begin{figure}[tbp]
\epsfxsize=8cm
\epsffile{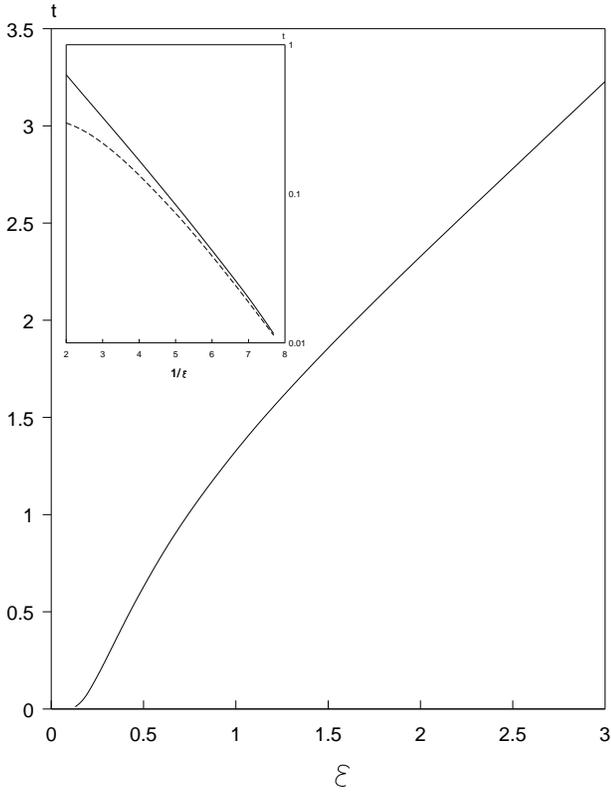}
\vspace{1cm}
\narrowtext
\caption{The conductivity exponent $t$ as a function of $\epsilon =d-1.$ The
  insert shows the comparison of numerical (solid curve) and analytical
  results.}
\label{numer}
\end{figure}

At arbitrary values of $\epsilon $ one can solve Eq. (\ref{main}) only
numerically. Numerical results concerning this solution under the threshold
boundary conditions: $Q\mid _{s=0}=1-c_t$, $Q\mid _{s=+\infty }=0$, support
the following assertion: the function $Q(s,\lambda )$ eventually evolves at
sufficiently large $ \lambda $ into the automodeling form:
\begin{eqnarray}
  Q(s,\lambda )= \overline{Q}(s\lambda ^a),~~~~~~ P(\sigma ,\lambda
  )=\overline{P}(\sigma \lambda ^{-a}),
\label{automodel}
\end{eqnarray}
where $a$ is the $\epsilon $--dependent exponent. This form ensures,
together with Eq. (\ref{rb}), the power--law critical behavior of the
conductivity $\Sigma $:
\begin{equation}
\Sigma(\tau) \sim \tau^t,~~~~~~~ \tau= {c-c_t\over c_t},~~~~~~~~t=a\nu.
\label{deft}
\end{equation}
Here $\nu $ is the critical exponent of the correlation length (the
characteristic size of a finite connected cluster) that is determined
from Eq. (\ref{rb}) by linearizing near the threshold point:
\begin{equation}
\nu^{-1}=\epsilon (\ln c_t+1)+1+\ln (1-c_t),
\end{equation}
where $c_t$ is given by Eq. (\ref{thres}). The critical index for the
conductivity, $t$, as a function of $\epsilon$ can be
found only numerically, and the result is shown in Fig. \ref{numer}.

\vskip 0.5cm

In the limit of small $\epsilon $ the problem may be treated analytically.  In
this case the threshold concentration and the critical index of the
correlation length are given by:
\begin{equation}
  c_t=e^{-1/\epsilon },~~~~~~ \nu ^{-1}=\epsilon -\frac 12e^{-1/\epsilon
    }+\ldots~.  \label{length}
\end{equation}
One can see, that, though the main part of $\nu ^{-1}(\epsilon )$ dependence
is linear, the remaining dependence on $\epsilon $ is essentially
nonanalytic, and therefore no regular $\epsilon $--expansion in $1+\epsilon $
dimensions is possible.

Let us look for the automodeling solution (\ref{automodel}) at $\epsilon \ll
1$. Using the fact that for $\epsilon =0$ this solution is simply $\exp (-s)$
($a=0$), we assume that
\begin{equation}
\overline{Q}(x)=(1-c_t)\exp [-x+v(x)],  \label{intv}
\end{equation}
as well as that $a\sim c_t$, $v\sim c_t$.  Plugging Eqs.
(\ref{intv},\ref{automodel}) and (\ref{length}) into Eq. (\ref{main}),
linearizing in $v$, and neglecting higher--order terms over $c_t$, such as
$av$, $c_t^2$, $c_tv$, etc., we arrive at the following equation:
\begin{equation}
\xi v^{\prime \prime }-\xi v^{\prime }+v=e^{-1/\epsilon }\left[ 1-\frac{J_1(2%
\sqrt{\xi })}{\sqrt{\xi }}\exp \left( \frac{\epsilon \xi }{1+\epsilon }%
\right) \right] -a\epsilon \xi \,,  \label{final}
\end{equation}
where $x=\epsilon \xi /(1+\epsilon )$.

The solution of Eq. (\ref{final}) with $v(0)=v^{\prime }(0)=0$ can be found by
the substitution: $v(\xi) = \xi w(\xi)$.  We have:
\begin{eqnarray}  \label{solution}
  w^{\prime }(\xi )&=&\frac{e^\xi }{\xi ^2}\left\{ e^{-1/\epsilon }[ 1-e^{-\xi
    }-\int\limits_0^\xi \frac{d\eta }{\sqrt{\eta }}e^{-\eta /(1+\epsilon
    )}J_1(2\sqrt{\eta })]\right. \nonumber\\ &~&~~~~~~\left.-\epsilon a\left[
  1-\left( 1+\xi \right) e^{-\xi }\right] \right\}.
\end{eqnarray}
To eliminate here the term, growing at $\xi \gg 1$ as $\exp \xi \simeq \exp
(x/\epsilon )$, the following condition should be fulfilled:
\begin{equation}
\epsilon a=c_t\left[ 1-\int\limits_0^\infty \frac{d\eta }{\sqrt{\eta }}%
e^{-\eta /(1+\epsilon )}J_1(2\sqrt{\eta })\right] .
\label{index1}
\end{equation}
This equation determines the critical exponent for the conductivity $t=a\nu $
(see Eq. (\ref{deft})). By combining Eqs. (\ref{length}) and (\ref{index1}) we
get:
\begin{equation}
t=\frac 1{\epsilon ^2}\exp \left( -1-\frac 1\epsilon \right) =\frac 1{\left(
d-1\right) ^2}\exp \left( -\frac d{d-1}\right) \,.  \label{index}
\end{equation}

The coefficients $v_n$ of a Tailor expansion of $v(x)$ at $x=0$ represent the
cumulants of the distribution function for resitivities, e.g. $v_2=\left(
\left\langle R^2\right\rangle -\left\langle R\right\rangle ^2\right)
  /\left\langle R\right\rangle ^2$. They are obviously of the order of $ c_t$.
  That means that the resistivity distribution at length scale $\lambda $ is
  of an almost gaussian $\delta $--function, centered at $\left\langle
  R\right\rangle \propto \lambda ^a$ and with the width
  $\propto\lambda^a~\exp(-1/\epsilon) $.

Using the scaling arguments \cite{s94,nyo94} one can derive the low--frequency
part of the AC conductivity $\Sigma \left( \omega ,\tau \right) $ near the
percolation threshold, if the critical exponent of the DC conductivity is
known. System's conductivity at low frequencies $\omega \ll W_0$ ($W_0$
is the hopping rate of the retaining bonds) and near the percolation
threshold, $\tau \ll 1$, is supposed to be represented in the scaling form
\begin{equation}
\Sigma (\omega ,\tau )=\frac{e^2n_e}{kT}a_{\parallel }^2|\tau |^tW_0g\left[
\tau \left( \frac S{W_0}\right) ^{-u/t}\right] ,  \label{scaling}
\end{equation}
where $S=-i\omega $, $a_{\parallel }$ is the longitudal lattice constant, and
\begin{equation}
u=\frac t{s+t}\;,\;\;\;\;\;s=2\nu -\beta .  \label{frexp}
\end{equation}
Here $s$ and $u$ represent the critical exponents for the dielectric constant
near the percolation threshold and the frequency-dependent conductivity at the
threshold, resp., and $\beta $ is the exponent of the percolation order
parameter (the probability of the site to belong to the infinite cluster). The
scaling function $g(x)$ has the following asymptotic behavior:
\[
g(x)\approx \left\{
\begin{array}{l}
A\left| x\right| ^{-t},\text{~as }\left| x\right| \ll 1\,; \\
1+B_{+}x^{-s-t}+\ldots, \text{~as }x\gg 1\,; \\
B_{-}\left| x\right| ^{-s-t}+\ldots, \text{~as }x<0,\;|x|\gg 1.
\end{array}
\right.
\]
The critical exponent $\beta $ in nearly one--dimensional systems may be
obtained to be:
\[
\beta =\frac 1{3\epsilon }c_t^2=\frac 1{3\epsilon }\exp \left( -\frac
2\epsilon \right) \,.
\]
It may be set equally to zero in all the following expressions due to a
sufficiently high order smallness of $c_t$, so that we have: $s\approx 2\nu
\approx 2/\epsilon $.

Now let us consider a nearly one--dimensional system with variable range
hopping along the chains. We assume \cite{md79} that the hopping rates depends
on the intersite distances $r_{ij}=\left| r_i-r_j\right| $ and on the energy
difference $\varepsilon _{ij}=\left| \varepsilon _i-\varepsilon _j\right| $
via $f_{ij}=r_{ij}/a+\varepsilon _{ij}/2kT$ ($a$ is the localization radius)
as: $W_{ij}=\omega _0\exp \left( -2f_{ij}\right) $.  Assuming site positions
and energies near the Fermi level homogeneously distributed  with the
density $N_F$, we arrive at the ``two--dimensional Poisson'' distribution for
$f_{ij}$, $F(f)=\exp \left[ -\left( f/f_0\right) ^2\right] $, with $f_0=\left(
T_0/T\right) ^{1/2}$, $kT_0=\left( N_Fa\right) ^{-1}$.

It is known \cite{bb85}, that the behavior of the conductivity  with
continuously distributed hopping rates may be explained, at least
qualitatively, if the result for the percolation lattice is known. For this
purpose, we choose some trial value of the hopping rates $W_c$. Then, the
initial system is replaced by the percolation network where the hopping rates
less than $ W_c$ are set to zero (broken bonds), and all others are set equal
to $W_c$. The conductivity of such a system is obviously less than the initial
one. If the trial value of $W_c$ is chosen so that the conductivity of the
percolation system is maximal, one can hope that this value gives a good
estimate for the conductivity of the real system.

Performing this procedure, e.g. at zero frequency, one can find, that the DC
conductivity is determined by the hops with $f_{ij}$ very close to its
threshold value $f_t$, which is determined by $F(f_t)=c_t$. From Eq. (\ref
{length}) it follows that $f_t=f_0/\sqrt{\epsilon }=\left( T_1/T\right)
^{1/2}$, where $T_1=T_0/\epsilon =\left( \epsilon N_Fa\right) ^{-1}$.
Taking in Eq. (\ref{scaling}) $n_e=N_FkT$ and $a_{\parallel }=1/n_e$, we
have:
\begin{equation}
\Sigma _{DC}=\frac{e^2 W_t}{N_F\left( kT\right) ^2}=\frac{e^2\omega_0}{%
N_F\left( kT\right) ^2}\exp \left[ -\left( T_1/T\right) ^{1/2}\right].
\label{dc}
\end{equation}
At the lowest frequencies (hydrodynamical region, $\omega \ll \omega _h$) the
conductivity can be close to its DC value and may be represented as a power
series in $-i\omega $: $\Sigma \left( \omega \right) =\Sigma _{DC}\left(
1-i\omega /\omega _h-\ldots \right) $, with a very small value for $\omega
_h$. As a result, the static dielectric constant $\eta$ becomes very
large and strongly temperature-dependent
\begin{eqnarray}
\eta \sim W_t/ \omega_h \sim (T_1/T)^{1/\epsilon} \exp(2/\epsilon^2)~.
\label{dconst}
\end{eqnarray}
At $\omega \gg \omega _h$ the frequency dependence of
conductivity  is given by:
\begin{equation}
\Sigma \left( \omega \right) =\frac{e^2 W_t}{N_F\left( kT\right) ^2}\exp
\left[ \epsilon \left( \frac{T_1}T\right) ^{1/2}\left( \frac{-i\omega }{W_t}%
\right) ^{\epsilon /2}\right].  \label{imed}
\end{equation}
Here, the temperature dependencies of the conductivity and of the dielectric
constant remain essentially of the ``1-d Mott's'' type. If the temperature is
relatively high, i.e. $T\gg T_2$ ($T_2=\epsilon ^4T_1/4$), this dependence
would continuously turn into one that is typical for $d=1$ (see, e.g.
\cite{nps89}) at frequencies of the order of $\omega _m$, $\ln ($ $\omega
_m/W_t)\sim -2/\epsilon $. At lower temperatures, $T\ll T_2$, Eq. (\ref{imed}%
) is valid only until $\omega \ll \widetilde{\omega }_m$, $\ln (\tilde \omega
_m/W_t)\sim -(1/\epsilon )\ln (T_2/T)$. At $\widetilde{\omega }_m\ll \omega
\ll \omega _M$, where $\ln(\omega_M/\widetilde{\omega}_m) \sim \epsilon
(T_1/T)^{1/2}$, we have the following dependence:
\begin{equation}
\Sigma \left( \omega \right) =\frac{e^2}{N_F\left( kT\right) ^2}\left(
-i\omega \right) \left[ \frac{\epsilon \left( T_1/T\right) ^{1/2}}{\ln
\left( -i\omega /\widetilde{\omega }_m\right) }\right] ^{2/\epsilon }\,,
\label{lttrans}
\end{equation}
which changes into one--dimensional one  (some power of temperature)  at $
\omega \sim \omega _M$.

In sum, by taking into account the specific fractal structure of polymer's
network we obtained the very specific temperature dependence for the static
conductivity (Eq.(\ref{dc})), dielectric constant (Eq.(\ref{dconst})) and
frequency dependent conductivity (Eqs.(\ref{imed},\ref{lttrans})).  The source
of this dependence is the following. The low dimensional random system can be
separated in weakly coupled clusters within which carriers occur to be
locked.  With temperature the size of clusters exponentially increases, since
more space becomes accessible for carriers due to thermal activation. As a
result the dielectric constant and the conductivity exhibit strong temperature
dependencies. There is also the very strong frequence dispersion of
conductivity starting with extremely low frequencies.  Such a type of strong
temperature and frequency dependence is actually observed in conducting
polymers wth localized carriers~\cite{wjrme90,joea94}.

Let us stress again that a similar dependence can not be obtained within the
standard $2d$ or $3d$ hopping models. By assuming the strong
Coulomb repulsion between carriers, the ``1d Mott's law'' could be
reproduced~\cite {shef84}, but it does not lead to any strong dependence for
the dielectric constant and for the frequency-dependent conductivity. The
reason is the following. In contrast to the low-dimensional case the clusters
in $2d$ and $3d$ systems prove to be more effectively coupled.  Therefore, the
large polarization of clusters does not happen because of leakage of carriers.
Thus, our results support strongly the idea that the conducting polymers
represent a low-dimensional substance even in the dielectric phase.

The authors are grateful to A. Epstein, G. Du, P. Fulde, M. Kastner, and B.
Shalaev for their interest in the problem. The work was partially supported by
a Czech National Grant GACR 202/94/0453.

\end{multicols}

\end{document}